\documentclass[aps,twocolumn,tightenlines,amsmath,amssymb,superscriptaddress]{revtex4-2}

\usepackage{graphicx}
\usepackage{amsmath}
\usepackage{amssymb}
\usepackage{epstopdf}
\usepackage{color}
\usepackage{lipsum}

\usepackage[colorlinks=true]{hyperref}
\hypersetup{citecolor=blue}

\usepackage{ulem}
\usepackage[dvipsnames]{xcolor}
\usepackage{todonotes}
\DeclareGraphicsExtensions{.eps}

\begin{document}
\title{Programmable multi-photon quantum interference in a single spatial mode}

\author{L.~Carosini}
	\email{lorenzo.carosini@univie.ac.at}
	\affiliation{University of Vienna, Faculty of Physics, Vienna Center for Quantum Science and Technology (VCQ), 1090 Vienna, Austria}	
	\affiliation{Christian Doppler Laboratory for Photonic Quantum Computer, Faculty of Physics, University of Vienna, 1090 Vienna, Austria}
\author{V.~Oddi}
	\affiliation{Dipartimento di Fisica, Politecnico di Milano, Piazza Leonardo da Vinci, 32, I-20133 Milano, Italy}
\author{F.~Giorgino}
	\affiliation{University of Vienna, Faculty of Physics, Vienna Center for Quantum Science and Technology (VCQ), 1090 Vienna, Austria}
	\affiliation{Christian Doppler Laboratory for Photonic Quantum Computer, Faculty of Physics, University of Vienna, 1090 Vienna, Austria}		
\author{L.~M.~Hansen}
	\affiliation{University of Vienna, Faculty of Physics, Vienna Center for Quantum Science and Technology (VCQ), 1090 Vienna, Austria}
	\affiliation{Christian Doppler Laboratory for Photonic Quantum Computer, Faculty of Physics, University of Vienna, 1090 Vienna, Austria}
\author{B.~Seron}
	\affiliation{Quantum Information and Communication, Ecole polytechnique de Bruxelles, CP 165/59, Universit\'e libre de Bruxelles (ULB), 1050 Brussels, Belgium} 
\author{S.~Piacentini}
	\affiliation{Dipartimento di Fisica, Politecnico di Milano, Piazza Leonardo da Vinci, 32, I-20133 Milano, Italy}
	\affiliation{Istituto di Fotonica e Nanotecnologie, Consiglio Nazionale delle Ricerche (IFN-CNR), Piazza Leonardo da Vinci, 32, I-20133 Milano, Italy}	
\author{T.~Guggemos}
	\affiliation{University of Vienna, Faculty of Physics, Vienna Center for Quantum Science and Technology (VCQ), 1090 Vienna, Austria}	
	\affiliation{Christian Doppler Laboratory for Photonic Quantum Computer, Faculty of Physics, University of Vienna, 1090 Vienna, Austria}		
\author{I.~Agresti}
	\affiliation{University of Vienna, Faculty of Physics, Vienna Center for Quantum Science and Technology (VCQ), 1090 Vienna, Austria}				
\author{J.~C.~Loredo}
	\email{juan.loredo@univie.ac.at}
	\affiliation{University of Vienna, Faculty of Physics, Vienna Center for Quantum Science and Technology (VCQ), 1090 Vienna, Austria}	
	\affiliation{Christian Doppler Laboratory for Photonic Quantum Computer, Faculty of Physics, University of Vienna, 1090 Vienna, Austria}
\author{P.~Walther}
	\affiliation{University of Vienna, Faculty of Physics, Vienna Center for Quantum Science and Technology (VCQ), 1090 Vienna, Austria}	
	\affiliation{Christian Doppler Laboratory for Photonic Quantum Computer, Faculty of Physics, University of Vienna, 1090 Vienna, Austria}

\begin{abstract}

The interference of non-classical states of light enables quantum-enhanced applications reaching from metrology to computation. Most commonly, the polarisation or spatial location of single photons are used as addressable degrees-of-freedom for turning these applications into praxis. However, the scale-up for the processing of a large number of photons of such architectures is very resource demanding due to the rapidily increasing number of components, such as optical elements, photon sources and detectors. Here we demonstrate a resource-efficient architecture for multi-photon processing based on time-bin encoding in a single spatial mode. We employ an efficient quantum dot single-photon source, and a fast programmable time-bin interferometer, to observe the interference of up to 8 photons in 16 modes, all recorded only with one detector––thus considerably reducing the physical overhead previously needed for achieving equivalent tasks. Our results can form the basis for a future universal photonics quantum processor operating in a single spatial mode.
\end{abstract}

\maketitle

Multi-photon interference lies at the heart of many optical quantum technologies. An optical quantum computer~\cite{Flamini_2018,PQIP_Pryde_2019} itself is in essence a large photonic multimode interferometer producing outcomes that can not be efficiently obtained by an otherwise classical device. Here, typically one starts by preparing as many single-photons as possible in different spatial modes, which are then fed into bulk-based~\cite{BS_Pan17}, or integrated~\cite{MultiD_Thompson_2018,Taballione_2021} interferometric networks. The most complex multi-photon experiments thus far have prepared sources of tens of single-photons~\cite{12photon_2018,BS_20photon_2019}, as well as squeezed states of light~\cite{QAdv_photon20,BS_Squeezed_Xanadu21}, interfering in circuits with more than 100 modes and detectors~\cite{BS_squeezed_2021}. While these are impressive achievements, it is clear that approaches of this kind are incredibly resource-demanding, simply requiring the control of unfeasibly many elements simultaneously––from large numbers of phase-locked sources, to high-voltage active electro-optical components, and costly superconducting nanowire single-photon detectors, to name a few. It is thus essential to develop methods that can still produce equivalent non-classical statistics, but efficiently use the available physical resources.

The most advanced technologies for producing multi-photon sources to date are either based on probabilistic frequency-conversion processes in non-linear crystals~\cite{Spring:17,Graffitti:18}, or obtained deterministically from the spontaneous emission of atomic transitions~\cite{SPS:Xing16,sps:senellart17,UPPU:20,Tomm:2021aa} and subsequent time-space demultiplexing~\cite{lenzini:17,BS_Pan17}. The former can be run either below a pump threshold to produce heralded single-photons, but with efficiencies kept low to mitigate the effect of unwanted higher-order photon emission, or they can run above threshold to produce squeezed states instead, however requiring complex phase-locking systems. On the other hand, the latter technology can deterministically produce single-photons that are also efficiently collected, with state-of-the-art (fibre) source efficiencies beyond {50\%}. Starting from one such source, active demultiplexing enables the construction of multi-phon sources. However, this demultiplexing step may not even be necessary. A standard one-photon source already contains all the necessary single-photons––they are all in different temporal modes, only sharing the same spatial trajectory. Thus, should one have access to devices that can alter their time-bin photon statistics, multi-photon interference can entirely occur in one single spatial mode.

Here, we demonstrate the resource-efficient interference and processing of a considerable number of particles, carried out with a very limited number of physical resources: one single-photon source, one programmable loop interferometer, and one single-photon detector. That is, by combining a quantum dot based photon source, from which we measure single-photon count rates at 17.1~MHz, together with a low-loss fast reconfigurable optical processor, and one highly efficient superconducting nanowire single-photon detector, we observe the interference of up to 8 photons in 16 modes, where all the multi-photon processing is carried out by analysing the time-tags of a single detector. Extensions of our results can enable a future resource-efficient universal quantum photonics processor.

	\begin{figure*}[htp!]
		\centering
		\includegraphics[width=.98\textwidth]{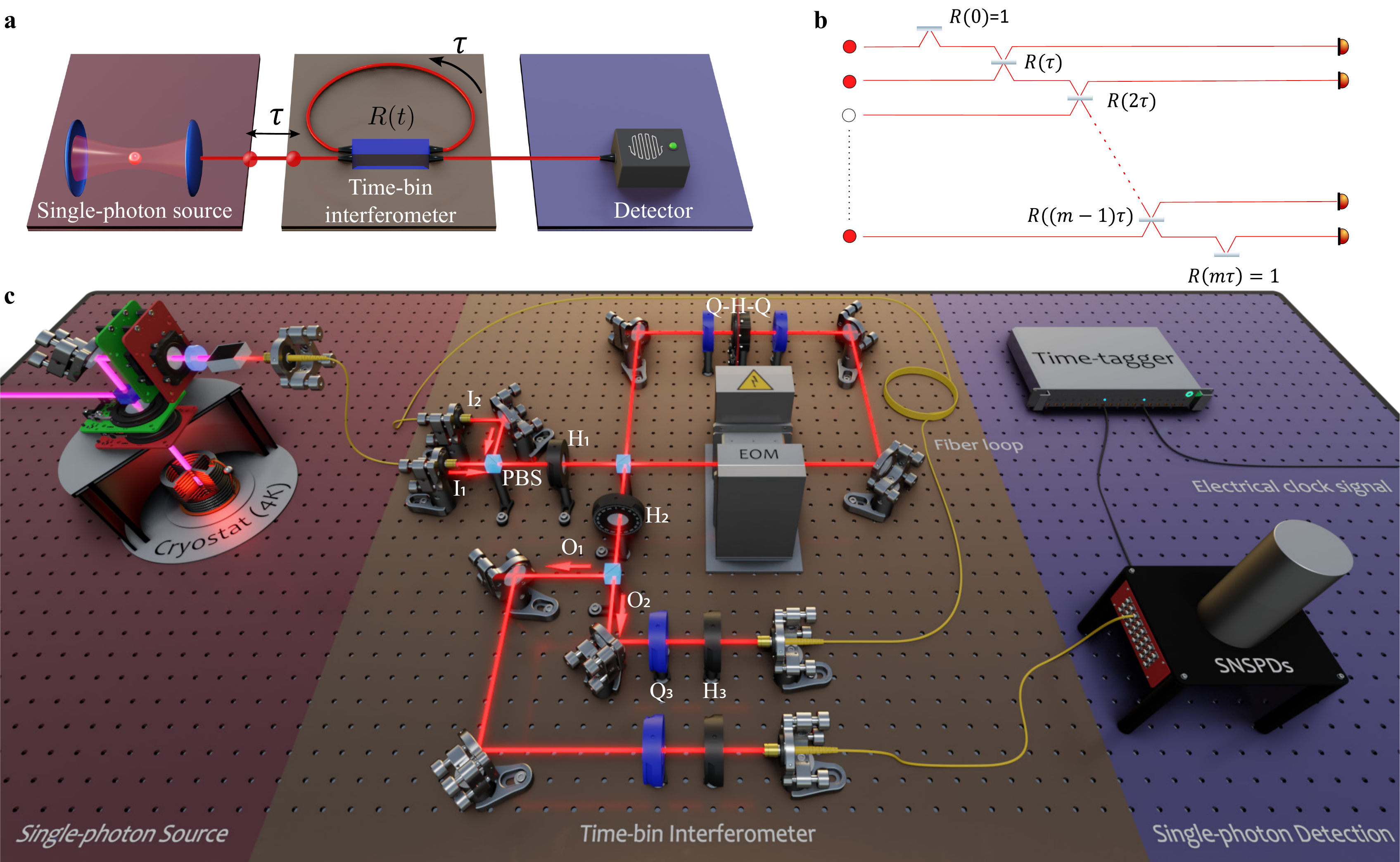}\vspace{0mm}	
		\caption{{\bf Time-bin multi-photon interferometric network.} {\bf a} Time-bin multiphoton processor. One single-photon source, one time-bin loop interferometer, and one detector implement the processor. {\bf b} Representation of time-bin multimode interferometer. The combination of a tuneable beamsplitter and delay loop implements a network of $m$ modes: an arbitrary sequence of beamsplitter operations between consecutive time-bins, with $m{-}1$ reflectivities $R(k\tau),~k{=}1,..,m{-}1$. The input time-bins contain either vacuum or a single-photon. Boundary initial and final reflectivities $R(0){=}R(m\tau){=}1$ correctly initialise and terminate the time-bin experiment. {\bf c} Depiction of the setup. The source (left) is an InGaAs quantum dot coupled to a micro-pillar cavity, kept at 4~K inside a cryostat. A confocal setup is used to pump and collect resonance-fluorescence, then sent to a time-bin interferometer: an effective time-varying beamsplitter with two inputs I$_1$, I$_2$ and two outputs O$_1$, O$_2$, with one output connected (looped) to one input via a $100$~ns fibre-based delay ($\approx 20$~m). The free-space electro-optic phase modulator (EOM) controls the time-varying reflectivity, which can be reconfigured to any value every $100$~ns. Half-wave plates H$_1$, and H$_2$, are kept at $\pi{/}8$ degrees. H$_3$, and quarter-wave plate Q$_3$ ensure that light traversing the loop arrives with vertical polarisation into loop input I$_2$ again. After traversing the loop a number of times, all photons and time-bins exit the interferometer and are detected with only one detector. The resulting statistics is reconstructed by post-processing events registered by the time-tagger.}
	\label{fig:1}
	\end{figure*}

\textit{Results.---}Figure~\ref{fig:1} describes the architecture we follow. It employs a time-bin interferometer composed of active and tuneable linear optical elements, and looped spatial trajectories, as proposed in Ref.~\cite{BS:Motes14}. The first step consists in triggering a single-photon source at time intervals $\tau$ to prepare a train of $n$ single-photons in $m$ designed time bins along a single spatial trajectory. The following step propagates the photon stream through a time-bin multimode interferometer, the core of which consists of a beamsplitter with time-varying reflectivity. Here, one output of the beamsplitter is connected back (looped) to one of its inputs, and traverses a delay matched to the arrival of a subsequent input photon after the time $\tau$. In this way, the device implements an arbitrary beamsplitter action between consecutive time-bins. This loop-based architecture is equivalent to a network with $m$ modes that are pairwise connected via beamsplitters with time-programmable reflectivities, as illustrated in Fig.~\ref{fig:1}b. We want to emphasize that such time-bin based scheme gives access to a universal liner-optics network when adding just a second phase-stable loop~\cite{BST:Rohde15}.

	\begin{figure*}[htp!]
		\centering
		\includegraphics[width=.8\textwidth]{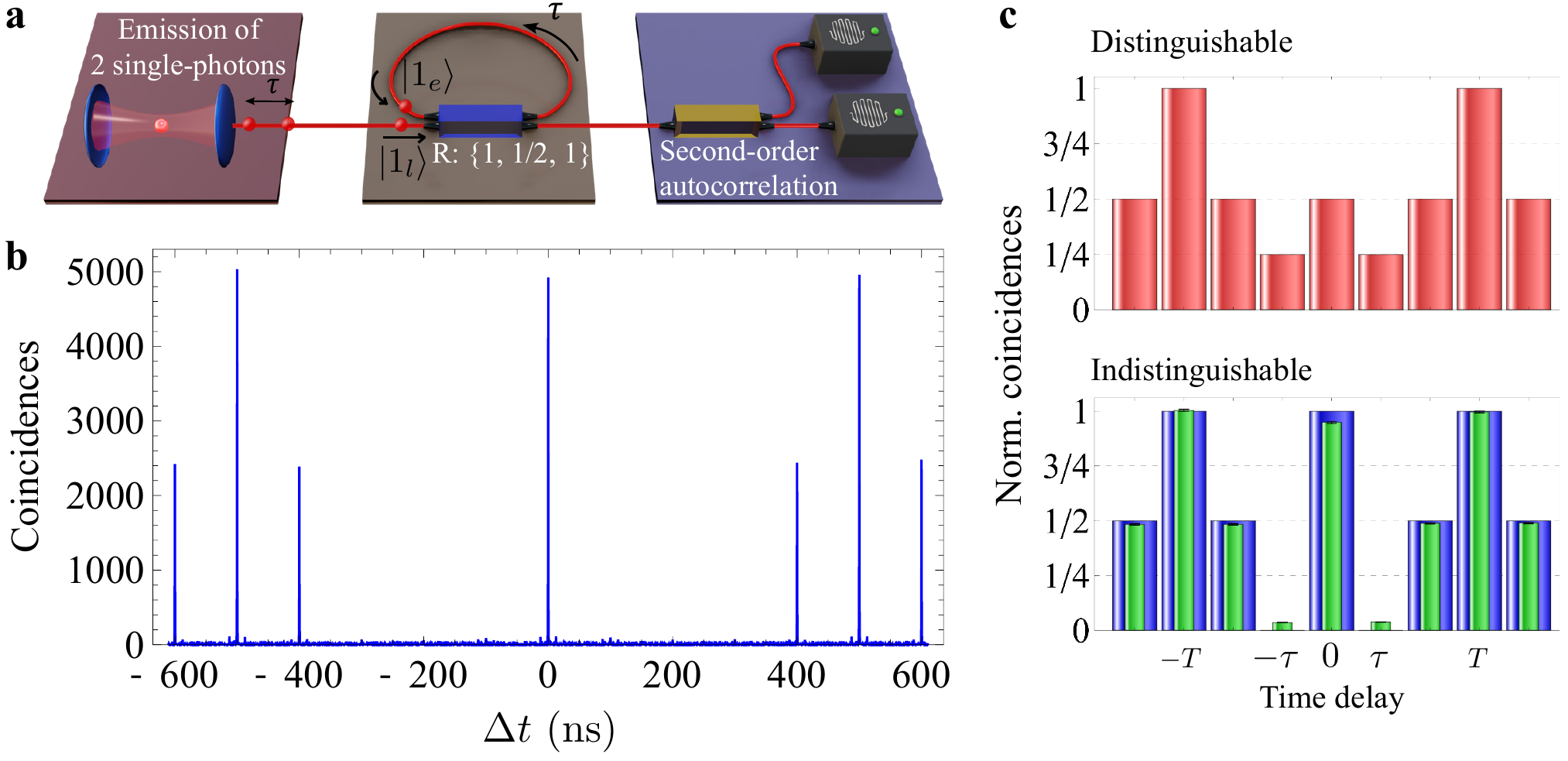}\vspace{0mm}	
		\caption{{\bf Time-bin HOM interference.} {\bf a} Protocol sequence. The source sends an early $|1_\text{e}\rangle$ and late $|1_\text{l}\rangle$ single-photon towards the time-varying looped interferometer. A sequence of reflectivities $R: \{1,1{/}2,1\}$ implements the time-bin HOM protocol. The output is analysed with a time-static beamsplitter and two detectors. {\bf b} Resulting second-order autocorrelation histogram. A series of coincidence peaks appear at different time delays. The signature of the time-bin HOM effect is the lack of coincidences at a delay equal to the photons' initial temporal separation $\tau{=}100$~ns. {\bf c} Normalised coincidences. Calculated for distinguishable photons input (red), indistinguishable photons input (blue), and extracted from our experimental data (green) by integrating the corresponding peak areas in a 3~ns window. The small error bars are estimated following Poissonian statistics of the detected events.}
	\label{fig:2}
	\end{figure*}

One main experimental challenge here is finding a physical implementation of an active looped interferometer such that it is fast-reconfigurable, low-loss, and capable of modifying and measuring the time-bin statistics entirely in a single spatial mode. Note that previous implementations of loop-based architectures~\cite{BST:He17,Madsen:2022aa,BST:Sempere22} have not yet met these conditions simultaneously. We go beyond previous experiments and demonstrate that these conditions can be fulfilled, see Fig.~\ref{fig:1}c. Our time-varying beamsplitter is built from a polarising beamsplitter-based Sagnac interferometer containing a free-space electro-optical phase modulator (phase-EOM) with a phase that can be reconfigured to an arbitrary value $[-\pi, \pi]$ every 100~ns. The polarising beamsplitters (PBSs) used here have throughput losses below 1\%, and polarisation extinction of more than 2000:1 for both output ports. The loss of the (anti-reflection coated) phase-EOM is also below 1\%. This allows building a fast-reconfigurable (10 MHz), high-visibility (0.998), and low-loss interferometer acting as our time-programmable beamsplitter. The output of this device is looped to one of its inputs and undergoes an optical delay of 100 ns, such that two propagating time-bins are always simultaneously arriving to the input ports of the tuneable beamsplitter. The fibre-based delay has an optical transmission of 94\%, mainly originating from the non-unity fibre-coupling efficiency.

Our single-photon source consists of resonance-fluorescence (RF) signal from a quantum dot-cavity device in a sample acquired commercially from Quandela. Laser pulses with $80$~MHz repetition rate coherently drive a trion transition––effectively, a two-level system at zero magnetic field. Our home-built optical setup is optimised to minimise collection losses, with a measured combined transmission of all free-space optical components of {$0.89{\pm}0.01$}. The single-photon free-space spatial mode is coupled to a single-mode fibre with a coupling efficiency of {$0.93{\pm}0.01$}, resulting in a first-lens to fibre polarised transmission of {$0.83{\pm}0.01$}. By driving our source at $\pi$-pulse excitation, we directly measure {17.1}~MHz of single-photons with a Single Quantum Eos superconducting nanowire single-photon detector (SNSPD) of {$0.85{\pm}0.02$} efficiency. At these conditions, we observe simultaneous high single-photon purity {$1{-}g^{(2)}(0){=}(98.61{\pm 0.01})\%$}, and indistinguishability {$I{=}(94.21{\pm 0.07})\%$}, see Supplementary Material.

To run our experiment, we chop pulses from the pump laser using a fibred electro-optical amplitude modulator (a-EOM), which upon excitation of the quantum dot prepares a train of $n$ single-photons in $m{\ge}n$ time-bins separated by $\tau{=}100$~ns. As a result, each time-bin is either occupied, or not, by a single-photon. After traversing the looped interferometer, the output photonic time-bin modes are sent into a SNSPD connected to a time-correlator---Time Tagger X, from Swabian Instruments---so that the output time-bin statistics can be revealed.
	
As a first test of our device, we performed a time-bin version of a Hong-Ou-Mandel (HOM) experiment~\cite{HOM87}. Here, we prepare our source such that two photons in consecutive time-bins, early and late, are sent to the loop interferometer. The active beamsplitter reflectivity, tuned via the phase of the EOM, is set to $R(0){=}1$ at the arrival of the first photon, hence it is sent deterministically into the loop. After $\tau{=}100$~ns, the reflectivity is set to $R(\tau){=}1{/}2$, so that the first and second photon interfere traversing two inputs of a balanced beamsplitter, see Fig.~\ref{fig:2}a. According to the rules that govern bosonic bunching, the two particles should leave the beamsplitter along the same output port, either escaping the loop or, with equal probability, both coupled into the loop again. After another $100$~ns, the reflectivity is programmed to $R(2\tau){=}1$, such that the time-bin within the loop exits entirely the setup and travels towards the detector. At the output of the time-bin interferometer, photon statistics is contained within one spatial mode, where in the ideal case both photons bunch at one of two consecutive time-bins, while terms with one photon in each time-bin are suppressed. Note that the time-bin before the first photon, as well as the one after the second photon must be empty for the correct operation of the protocol.

To help illustrate the result of this case, only for this measurement we use two SNSPDs and perform a standard second-order autocorrelation measurement at the time-bin interferometer output. Figure~\ref{fig:2}b displays the resulting coincidence histogram, and contains: (i) strong correlations at $\Delta t{=}0$, originating from the photon bunching terms---two photons occupying the same temporal mode lead to coincidences at zero delay after passing through a time-static beamsplitter---as well as (ii) suppressed correlations at $\Delta t{=}100$~ns---i.e., between consecutive time-bins---which is a signature of the time-bin Hong-Ou-Mandel effect. We evaluate the time-bin 2-photon interference visibility via $V^{(2)}{=}1{-}2C_{|\tau|}{/}\left(C_{|\tau|}{+}C_{0}\right)$, where $C_{|\tau|}{=}C_{\tau}{+}C_{-\tau}$ is the sum of areas of correlated peaks at $\Delta t{=}{\pm}\tau$, and $C_0$ is the area of the correlated peak at $\Delta t{=}0$. Normalisation implies $C_{\tau}{+}C_{-\tau}{+}C_0{=}C$, with $C$ the area of uncorrelated peaks at delays larger than the bins separation, located at $\Delta t{=}T{=}\pm500$~ns in Fig.~\ref{fig:2}b, with $T$ the period at which the experiment (sequence) is repeated. Note that for distinguishable particles, one expects $C_{\tau}{=}C_{-\tau}{=}C{/}4$, and $C_0{=}C{/}2$, whereas fully indistinguishable photons lead to $C_{\tau}{=}C_{-\tau}{=}0$, and $C_0{=}C$. Figure~\ref{fig:2}c displays these values of correlated and uncorrelated peaks (coincidences) normalised to the reference $C$, calculated for the case of fully distinguishable particles (red bars), the fully indistinguishable case (blue bars), together with the values obtained from our experiment (green bars). From our data we obtain $V^{(2)}{=}(85.97{\pm}0.06)\%$. Note that this value is affected by several factors, such as residual multi-photon emission from the quantum dot, photon distinguishability at increasing time-scales~\cite{loredo2016scalable,homTime:Pan}, imperfect active switching of the time-bin interferometer, as well as imperfect modulation of the laser pump stream that allocates either a single-photon or vacuum inside a time-bin.

One strong advantage of active time-bin interference is that the size of the implemented protocol can be increased, and programmed, without increasing the physical size of the experimental apparatus. For instance, the number of photons and modes is defined by choosing a pattern of time-bins that contain, or not, a single-photon, and the specific linear-optical network is set by specifying time-varying values of reflectivities. Hence, this protocol allows to straightforwardly increase the number of interfering photons $n$, where the practical limit is set by overall experimental efficiencies leading to an exponentially decreasing rate of $n-$photon events.

	\begin{figure}[htp!]
		\centering
		\includegraphics[width=.45\textwidth]{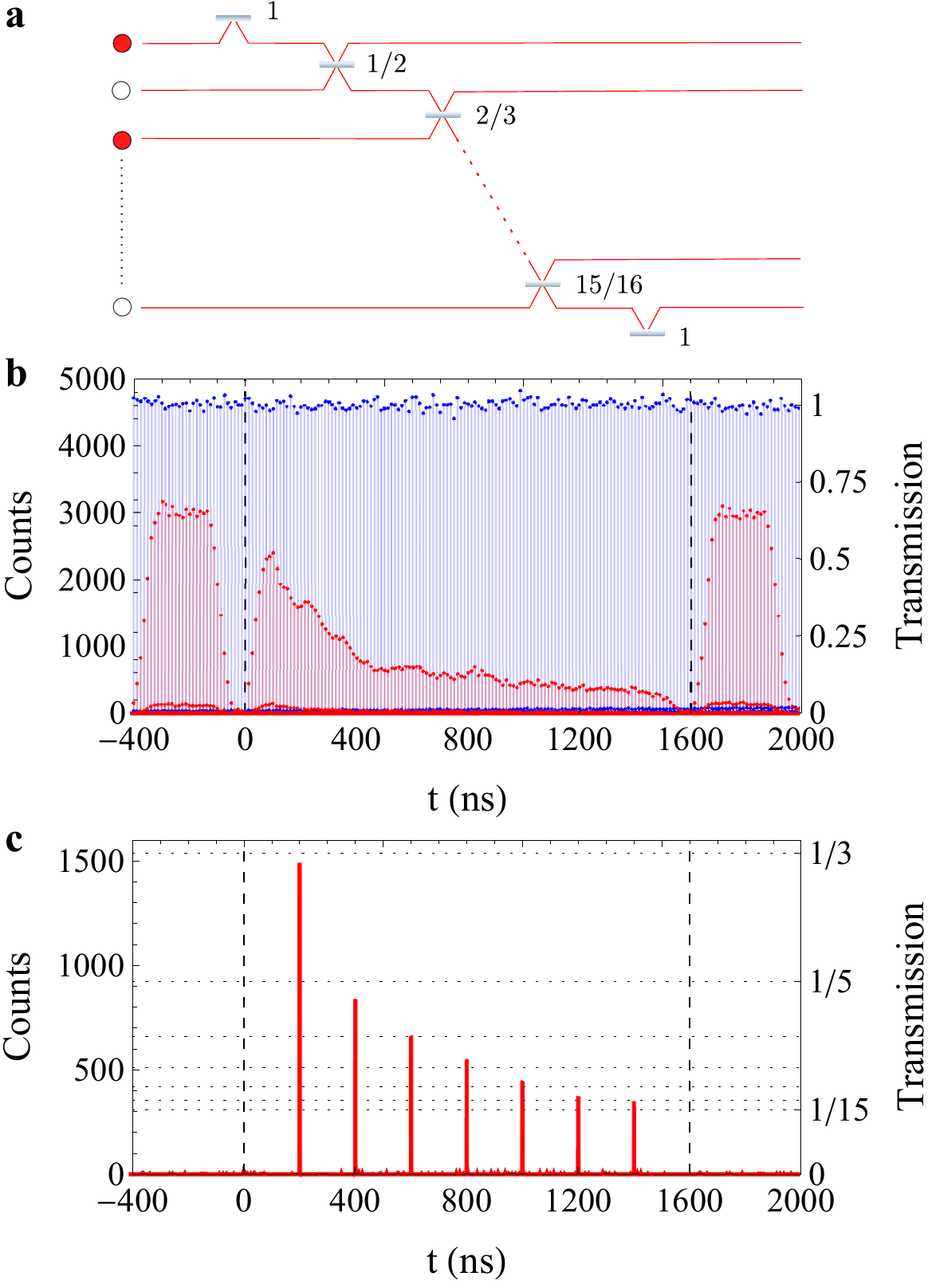}\vspace{0mm}	
		\caption{{\bf Time-varying network.} {\bf a} 8-photon network. $n{=}8$ single-photons occupy odd positions in $m{=}16$ modes, in a network with reflectivities $R_k{=}k{/}\left(k{+}1\right), k{=}1{,}{..}{,}15$. {\bf b} Measured time-dependent counts at the output of the time-bin interferometer. The loop is blocked, hence only the transmitted intensity is observed without (blue), and with (red) phase modulation. The area within the vertical dashed lines denotes the time lapse of interest, i.e., that containing 16 time-bins separated by 100~ns. The time duration outside this window can be chosen to take any value, in consequence defining the effective repetition rate of the experiment. Modulation of the laser pump (pulse chopping) is off, thus a peak appears every $12.5$~ns. All time-varying values of transmission $1{-}R_k$ occur between $t{=}0$, and $t{=}1.6$~$\mu$s. {\bf c} Pulse chopping on. Only 8 photons can now appear among all time-bins. Note that the first photon at $t{=}0$ is not visible, as transmission at this point is zero (boundary reflectivity $R(0){=}1$ sends the photon into the loop). This sequence of photons, synchronised with the sequence of reflectivities, perform the time-bin 8-photon protocol.}
	\label{fig:3}
	\end{figure}

	\begin{figure}[htp!]
		\centering
		\includegraphics[width=.45\textwidth]{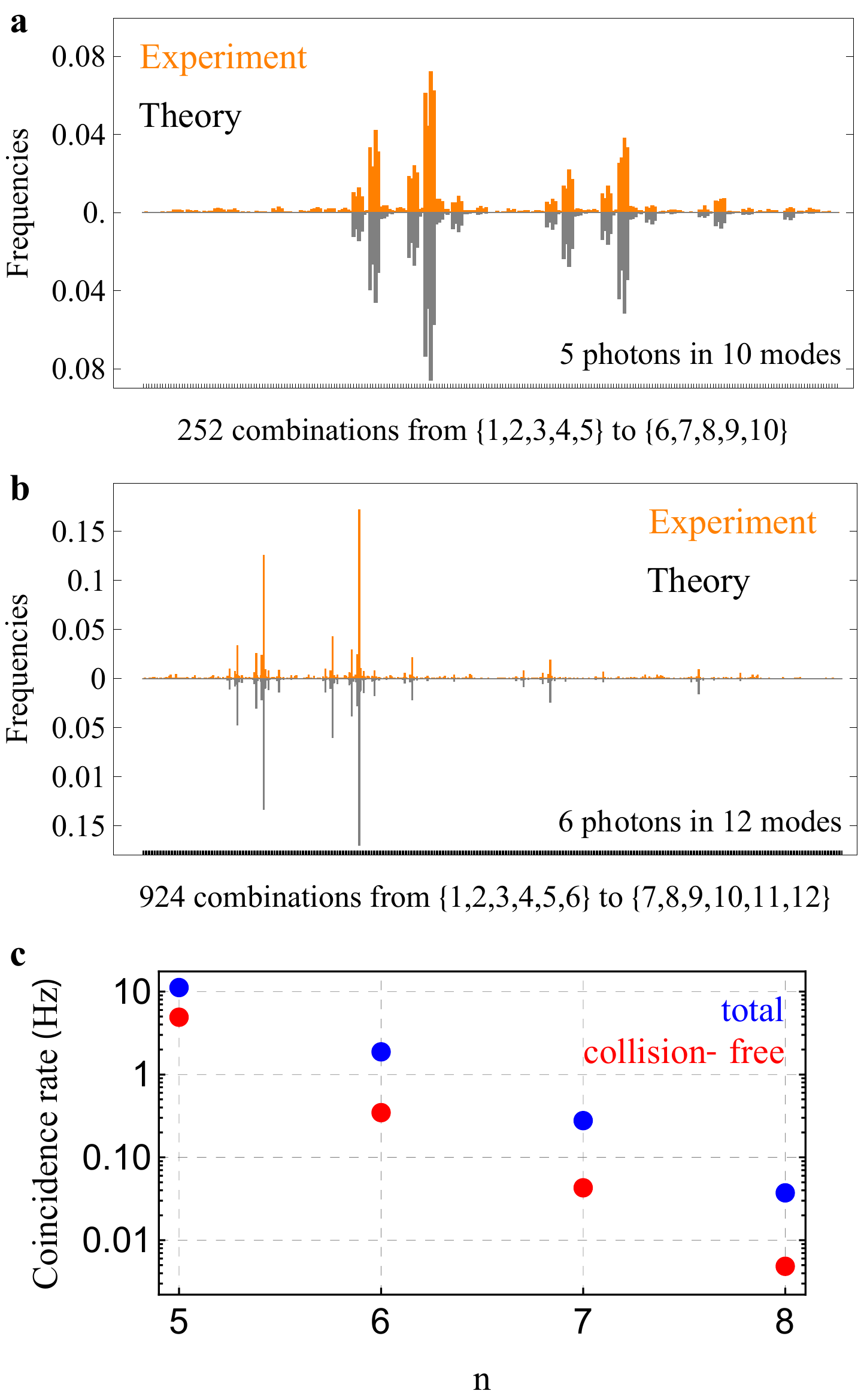}\vspace{0mm}	
		\caption{{\bf Reconstructed multi-photon output distribution.} Experiment (orange) and theory (grey) output frequencies for {\bf a} $n{=}5$ photons in $m{=}10$ modes, with reflectivities $R_k{=}\left(0.5,0.6,0.7,0.8,0.8,0.8,0.7,0.5,0.4\right)$, and photons occupying even input modes $\left\{2,4,6,8,10\right\}$; and {\bf b} for $n{=}6$, $m{=}12$, $R_k{=}k{/}\left(k{+}1\right), k{=}1{,}{..}{,}11$, and photons in odd inputs $\left\{1,3,5,7,9,11\right\}$. {\bf c} Coincidence rates. The measured rates (red dots) are only on the collision-free subspace. The total rates (blue dots) are estimated from dividing the measured ones by the accumulated probability of the collision-free subpsace.}
	\label{fig:4}
	\end{figure}

Accordingly, we now program our device to allow for the interference of a larger number of photons. In particular, we choose to place $n$ photons in $m{=}2n$ modes, for $n$ up to 8 single-photons---number constrained mainly by our source efficiency. In every experiment, the first input time-bin is deterministically routed into the loop, and the $m{+}1$ input time-bin is deterministically routed out from the loop. When the first and second time-bins interfere, some amplitude escapes the loop, defining the first output time-bin. The sequentially travelling photons keep interfering through the loop via the programmed beamsplitter operations, and then after a time $m{\times}\tau$, the last beamsplitter reflectivity is chosen such that the remaining amplitude in the loop deterministically leaves it, defining the $m{-}$th output time-bin. In order to showcase the programmability of our time-bin processor we targeted specific sets of reflectivities, in particular we choose $R_k{=}k{/}\left(k{+}1\right)$ for $k{=}1{,}{..}{,}m{-}1$, motivated by their application in, e.g., investigating the quantum central limit theorem~\cite{Becker:2021aa}. Fig.~\ref{fig:3} shows one example of these time-varying reflectivities and network, for an experiment with $n{=}8$ photons in $m{=}16$ modes. The reflectivity values are programmed by setting the voltage waveform produced by the arbitrary function generator that drives the phase-EOM. Note that in some cases, e.g., in Boson Sampling protocols, a number of modes scaling as $m{\sim}n^2$ is preferred. In our case, our scaling choice suffices as we no longer benefit from further increasing the number of modes given the specifics of the implemented network.

	\begin{figure*}[htp!]
		\centering
		\includegraphics[width=.98\textwidth]{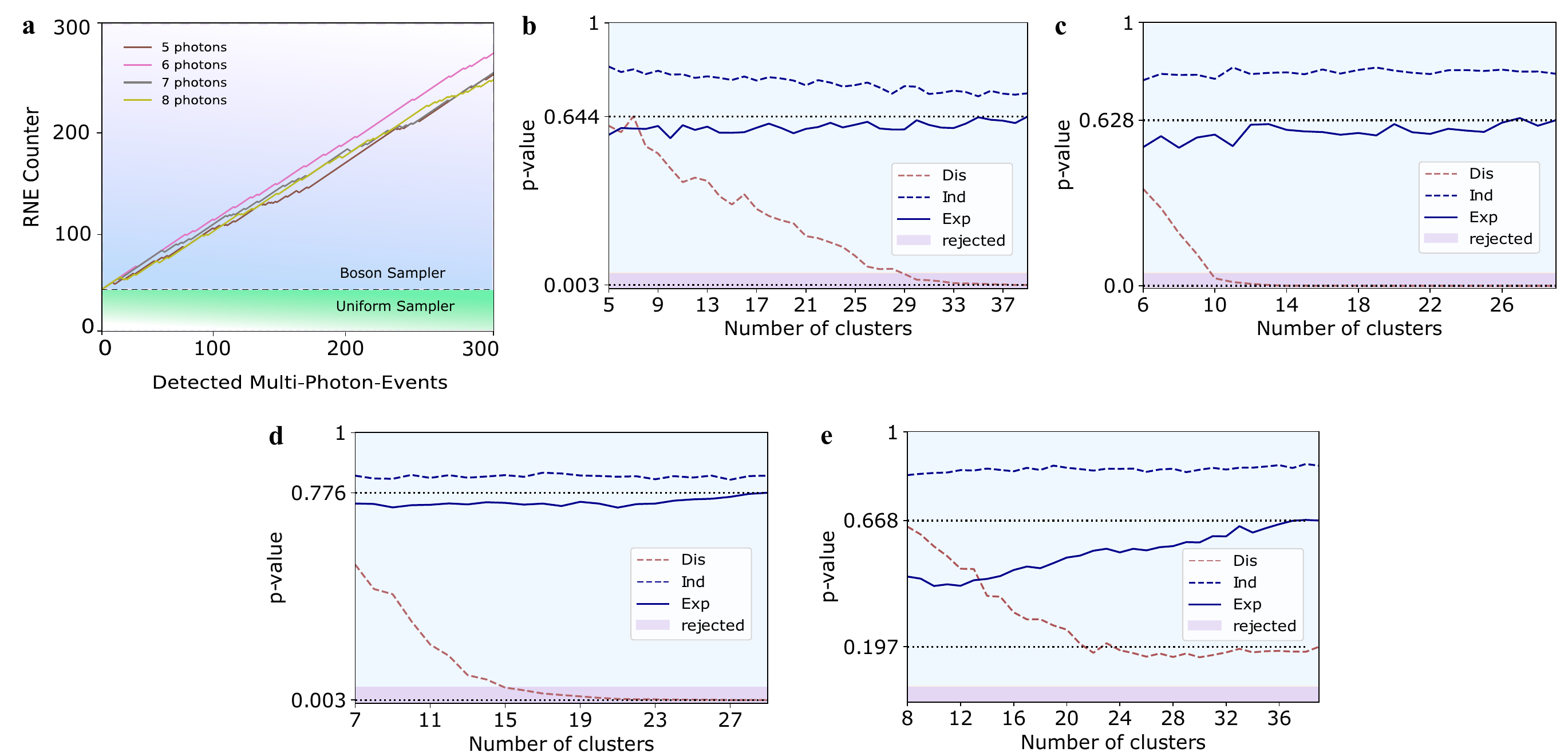}\vspace{0mm}
		\caption{{\bf Experimental validation.} {\bf a} Validation against uniform sampler. For every detected multi-photon event, a row-norm estimator (RNE) is calculated and a counter is updated according to its value. The increasing counter indicates that data is obtained from genuine bosonic interference. For clarity we included only the first 300 randomly chosen events for each experiment. {\bf b-e} Validation against distinguishable sampler for the experiments with 5, 6, 7, and 8 photons, respectively. The plots show the p-values obtained through the machine-learning based validation method employed to exclude that photons in our apparatus were not displaying bosonic interference. The dashed curves refer to numerically generated samples, which were used to select the hyperparameter values (namely number of clusters and sample size) ensuring that the algorithm could effectively discard samples drawn from the distinguishable-photon inputs and recognize as compatible those drawn from the indistinguishable-photon inputs. The size of the compared samples ranges from 300 events (for the 8-photons experiment) to 700 (for the 6-photons experiment). Whenever the size of the experimental samples was larger than the chosen sample size, we randomly extracted samples from the available experimental data 100 times and evaluated the p-value corresponding to the mean of the $\chi^2$ variables. Only for the case of $n=8$, the small sample-size is too low to conclusively reject samples drawn from the distinguishable-photon input distribution. However, we can observe that the experimental p-value converges closer to the indistinguishable sample case.}
	\label{fig:5}
	\end{figure*}

With these tools, we program multi-photon interference experiments with various numbers of photons. For example, Fig.~\ref{fig:4} displays instances with $5$, and $6$ photons, and distinct patterns of reflectivities. For each $n$-photon experiment, we sample the output time-bin distribution by measuring collision-free $n$-photon events, i.e, all those events in which $n$ photons output the interferometer in $n$ different output time-bin modes. The number of possible such events is given by $\binom{m}{n}$; by recording the amount of detected events for each of these combinations, we experimentally assess their relative frequencies, that is, output probabilities normalised within the collision-free subspace. Importantly, these measurements are carried out using only one SNSPD, and analysing the registered time-tags, see Supplementary Material. To obtain the expected output probabilities we model the involved optical circuits, including the effect of optical losses, with software available both in Python~\cite{Perceval:2023} and Julia~\cite{BosonJulia:2023}. The agreement between experiment and theory is quantified by the statistical fidelity $\mathcal{F}{=}\sum_i\sqrt{p^\text{exp}_ip^\text{th}_i}$ between their normalised distributions, or frequencies. We find $\mathcal{F}^{(5)}{=}0.939$, and $\mathcal{F}^{(6)}{=}0.914$ for the cases with $n{=}5$, and $n{=}6$ photons, respectively. Figure~\ref{fig:4}c displays the measured coincidence rates of our n-photon interference experiments. The rate of the collision-free subspace for 8 interfering photons is already at the 5~mHz level, at which point we stop adding particles.

To provide further evidence of bosonic quantum interference in our interferometer, we adopt two validation techniques against alternative explanations for our experimental statistics, which are widely employed for validating the Boson Sampling protocols on photonic platforms~\cite{BS_review_Brod}, see Supplementary Material for further details on simulations and validation techniques. We first exclude the hypothesis that the observed distribution arises from the uniform scattering of the samples~\cite{AAValidation}. Figure~\ref{fig:5}a depicts the results of this method for validating experiments with up to 8-photon interference. Indeed, the increasing positive counters with registered multi-photon events indicates that our data does not result from a uniform sampler. 

Furthermore, we apply a machine-learning-based validation method~\cite{agresti_2019} to rule out the hypothesis that our data originates from distinguishable photons input, hence one that displays no quantum interference. This technique is based on the comparison of two samples (a \textit{bona fide} and a test one) and estimates the probability that they are drawn from the same probability distribution. 
This is done by coarse graining---namely clustering---the \textit{bona fide} sample and then assigning the events of the test one to the created clusters. The employed clustering technique is the K-means method~\cite{macqueen1967classification, arthur2006k}. In the end, for each cluster, the number of events belonging to the two samples are compared through a $\chi^2$ test, whose p-value evaluates the probability that the two samples are drawn from the same distribution. A p-value $<0.05$ implies that distinguishable samples should be rejected. To validate our data, we numerically generate \textit{bona fide} samples, drawn from the expected distribution, corresponding to indistinguishable photon inputs, and use our observed frequencies as test sample. However, the effectiveness of this technique relies on the proper choice of its hyperparameters, e.g., sample size and number of clusters. Figure~\ref{fig:5}b shows the results of this method for the validation of our experiments with up to 8-photon interference. We take samples from the theoretical distributions derived from indistinguishable-photon inputs (compatible to the \textit{bona fide} samples), as well as others drawn from the alternative hypothesis of distinguishable-particle inputs (to be rejected). The results of the simulated protocols are displayed as dashed curves in Fig.~\ref{fig:5}b, following their expected behaviour. Notably, the experimental data (solid curves) closely follows the behaviour of indistinguishable photon input, hence supporting the hypothesis for bosonic quantum interference.
 
\textit{Discussion.---}We experimentally demonstrated time-bin reconfigurable multi-photon quantum interference occurring within one spatial mode. The size of the implemented experiments were chosen and increased, while maintaining a constant number of (programmable) physical resources: one single-photon source, one active time-bin processor, and one detector. We used validation protocols to certify that our data originates from quantum interference, in particular with a recently introduced machine-learning based approach~\cite{agresti_2019}.

In practice, the size and complexity of the experiment is constrained mainly by source efficiencies, in our case allowing to observe up to 8-photon interference. Note that when using only one loop in a time-bin experiment, the resulting collision-free n-photon rate also decreases for larger number of photons. This can be avoided by adding a second outer loop, as in Ref.~\cite{BS:Motes14}, to allow for full-connectivity in the optical circuit, in which case most $n$-photon events land in the collision-free subspace. Moreover, allowing for lost photons~\cite{BSwPL:16,BS_20photon_2019} is straightforward in our implementation, then increasing the rates of detected multi-photon events. However, in our current implementation we restrain from this approach, as it might rapidly lead to classical output distributions.

Our results show that the architecture is highly resource-efficient in comparison to the standard spatial-encoding approach, as it does not require neither active demultiplexing of a single-photon source, nor building arrays of multiple identical emitters. The feasibility of our architecture is underlined by observing tuneable multi-photon interference for, in principle, arbitrary number of particles and modes. We expect that future developments will aim to add a second outer phase-stable loop for achieving universal linear-optics quantum processing in a single-spatial mode.

\textit{\bf Funding.} This research was funded in whole, or in part, from the European Union’s Horizon 2020 and Horizon Europe research and innovation programme under grant agreement No 899368 (EPIQUS), the Marie Skłodowska-Curie grant agreement No 956071 (AppQInfo), and the QuantERA II Programme under Grant Agreement No 101017733 (PhoMemtor); from the Austrian Science Fund (FWF) through [F7113] (BeyondC), and [FG5] (Research Group 5); from the Austrian Federal Ministry for Digital and Economic Affairs, the National Foundation for Research, Technology and Development and the Christian Doppler Research Association. For the purpose of open access, the author has applied a CC BY public copyright licence to any Author Accepted Manuscript version arising from this submission.

\textit{\bf Acknowledgment.} The authors thank Patrik Zah\'alka for assistance with FPGA electronics and signal processing. 

\bibliography{biblio_loop}

\end{document}